\def \ts     {\thinspace}
\def \kms    {\ifmmode{{\rm \ts km\ts s}^{-1}}\else{\ts km\ts s$^{-1}$}\fi}
\def \msol   {\ifmmode{{\rm M}_{\odot} }\else{M$_{\odot}$}\fi}
\def \lsol   {\ifmmode{L_{\odot}}\else{$L_{\odot}$}\fi}
\def \lfir   {\ifmmode{L_{\rm FIR}}\else{$L_{\rm FIR}$}\fi}

\def \,{\thinspace}
\def \ppm{$\pm$}

\def \ppm{$\pm$}
\def \ly{Ly$\alpha$}
\def \arcsec{^{\prime\prime}}
\def \arcmin{^{\prime}}
\def \co{CO J\,=\,7$-$6}

\def\ctwo {\ifmmode{{\rm C}{\rm \small I}(^3\!P_2\!\to^3\!P_1)}
     \else{C\ts {\scriptsize I}{\small$(^3\!P_2\!\to^3\!\!\!P_1)$}}\fi}

\documentclass{naturesubmissionstyle}

\makeatletter
\renewcommand\@biblabel[1]{#1.}
\makeatother

\usepackage{graphics,graphicx}
\usepackage[super,comma]{natbib}
\usepackage[switch]{lineno}
\usepackage{authblk}
\usepackage{amssymb}
\usepackage{xcolor}
\usepackage{ulem}

\title{Evidence for Infalling Gas in a Lyman-$\alpha$ Blob}
\author{Yiping Ao$^{1,2,3}$, Zheng Zheng$^{4}$, Christian Henkel$^{5,6}$, 
Shiyu Nie$^{4}$, Alexandre Beelen$^{7}$, Renyue Cen$^{8}$, Mark Dijkstra$^9$,
Paul J. Francis$^{10}$, James E. Geach$^{11}$, 
Kotaro Kohno$^{12}$, Matthew D. Lehnert$^{13}$, Karl M. Menten$^{5}$, Junzhi
Wang$^{14}$, Axel Weiss$^{5}$}

\date{}

\begin{document}


\maketitle

\let\thefootnote\relax\footnote{
\begin{affiliations}
\item Purple Mountain Observatory \& Key Laboratory for Radio Astronomy,
	Chinese Academy of Sciences, 8 Yuanhua Road, Nanjing 210034, China,
	ypao@pmo.ac.cn
\item School of Astronomy and Space Science, University of Science and Technology of China, Hefei 230026, Anhui, China
\item National Astronomical Observatory of Japan, 2-21-1 Osawa, Mitaka, Tokyo
	181-8588, Japan
\item Department of Physics and Astronomy, University of Utah, Salt Lake City, UT 84112, USA
\item MPIfR, Auf dem H\"{u}gel 69, 53121 Bonn, Germany
\item Astron. Dept., King Abdulaziz Univ., P.O. Box 80203, Jeddah 21589, Saudi Arabia
\item Institut d’Astrophysique Spatiale, CNRS UMR 8617, Université Paris-Sud, Université Paris-Saclay 91405 Orsay, France
\item Department of Astrophysical Sciences, Princeton University, Princeton, NJ 08544, USA
\item Institute of Theoretical Astrophysics, University of Oslo, P.O. Box 1029 Blindern, NO-0315 Oslo, Norway
\item Physics Education Centre and Research School of Astronomy and Astrophysics, The Australian National University, Canberra ACT 0200, Australia
\item Centre for Astrophysics Research, School of Physics, Astronomy \& Mathematics, University of Hertfordshire, Hatfield, AL10 9AB, UK
\item Institute of Astronomy, The University of Tokyo, 2-21-1 Osawa, Mitaka, Tokyo 181-0015, Japan
\item Sorbonne Universit\'{e}, CNRS, UMR 7095, Institut d'Astrophysique de Paris, 98bis bd Arago, 75014 Paris, France 
\item Shanghai Astronomical Observatory, Chinese Academy of Sciences, 80 Nandan Road, Shanghai, 200030, China 
\end{affiliations}
}

{\noindent \bf Lyman-$\alpha$ blobs (LABs) are spatially extended nebulae of
emission in the \ly\, line of hydrogen, seen at high
redshifts\cite{Francis1996,Steidel2000}, and most commonly found in the dense
environment of star-forming galaxies\cite{Matsuda2004,Matsuda2012}.  The origin
of Ly$\alpha$ emission in the LABs is still unclear and under
debate\cite{Yajima2013}.  Proposed powering sources generally fall into two
categories: (1) photoionization, galactic super-winds/outflows, resonant
scattering of \ly\, photons from starbursts or active galactic nuclei
(AGNs)\cite{Colbert2006,Cen2013,Geach2016,Ao2015,Ao2017} and (2) cooling radiation
from cold streams of gas accreting onto galaxies\cite{Dijkstra2009}.
Here we analyse the gas kinematics within an LAB providing rare observational
evidence for infalling gas. This is consistent with the release of
gravitational accretion energy as cold streams radiate \ly\, photons.  It
also provides direct evidence for possible cold streams feeding the central
galaxies. The infalling gas is not important by mass but hints at more than one
mechanism to explain the origin of the extended \ly\, emission around young
galaxies.  It is also possible that the infalling gas may represent material
falling back to the galaxy from where it originated, forming a galactic
fountain.}

A recent study shows that nearly 100\% of the sky is covered by \ly\, emission
around high redshift galaxies\cite{Wisotzki2018,Laursen2019}.  Ionizing photons from young
stars in star-forming (SF) galaxies and/or unobscured AGNs can ionise neutral
hydrogen atoms in the circumgalactic medium and subsequent recombination leads
to \ly\, emission\cite{Cantalupo2017}. Recent studies have revealed sources of
radio or submillimeter dust emission within the
LABs\cite{Colbert2006,Ao2015,Ao2017}, suggesting that these embedded sources
are the main engines powering the extended \ly\, emission via
recombination radiation\cite{Cantalupo2017}.  Resonant scattering
of \ly\, photons in the circumgalactic medium leads to spatially extended
emission\cite{Cen2013,Colbert2006,Geach2009,Zheng2011}. Besides the
photoionisation mechanism, there is also supporting evidence for the cold
stream scenario from those LABs lacking any visible power
source\cite{Nilsson2006,Smith2007}. This cold stream mode has been further
demonstrated by recent simulations\cite{Dijkstra2009} --- if $>$10
per cent of the change in the gravitational binding energy of a cold flow goes
into heating of the gas, the simulated cold streams become spatially
extended \ly\, sources that are comparable to observed LABs. This model can
naturally explain the spatial distribution of the LABs and the diversity of
host galaxies in the LABs, as the \ly\, emission may be decoupled from the
associated central sources. The most luminous gravitationally powered blobs
would be associated with the most massive halos, which may host a variety of
sources like AGNs, Lyman Break Galaxies (LBGs) and Submillimeter Galaxies
(SMGs). However, both cold stream\cite{Dijkstra2009} and SF-based
models\cite{Cen2013} can reproduce the measured luminosity functions of LABs.
So far, there is a lack of direct observational evidence for cold stream
powered LABs.  Here we present observational evidence for the association of
\ly\, emission with gas infalling onto an LAB, 
suggesting cooling radiation as an alternative mechanism of 
\ly\, emission in some cases.

\begin{figure}[t]
\centering
\includegraphics[angle=0,width=0.9\textwidth]{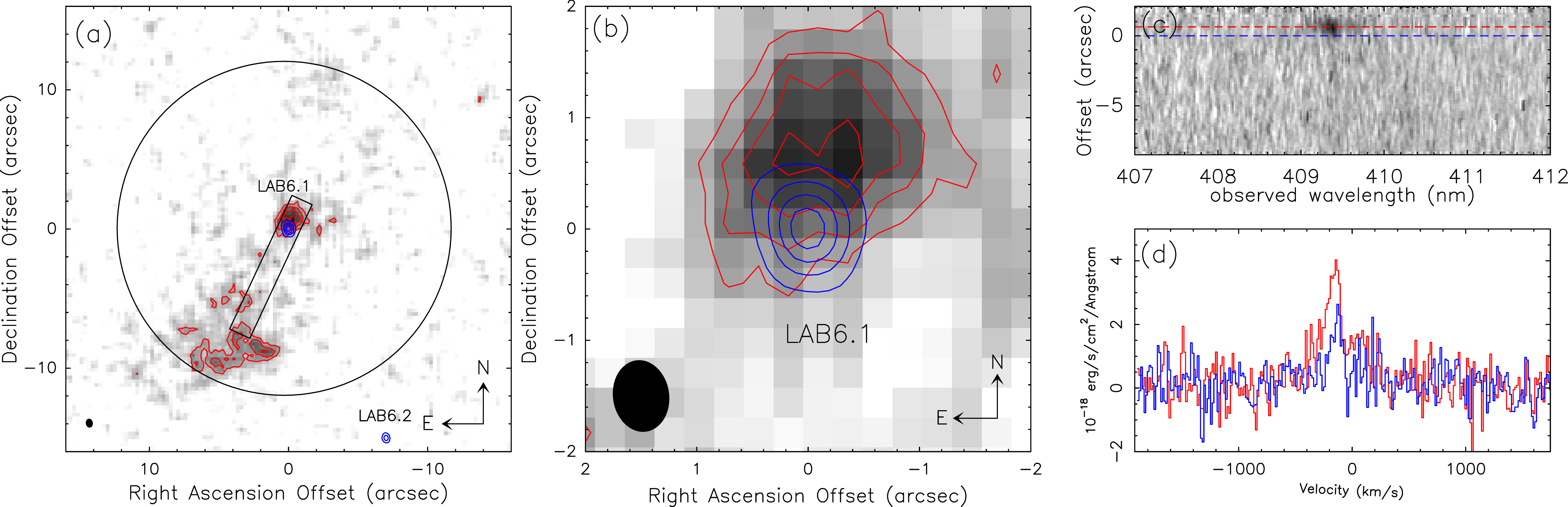}
\caption{{\bf Left:} The ALMA 1.2mm continuum image in blue contours (contour
	levels are 5, 10, 15, and 20$\sigma$ and 1$\sigma$\,=\,40~$\mu$Jy/beam)
	overlaid onto the \ly\, image\cite{Palunas2004}
	in grey (also shown as red contours with contour levels of 30\%,
	50\%, 70\% and 90\% of the peak intensity).  The two continuum sources
	detected at $>$5$\sigma$ levels are labeled.  The big circle shows the
	primary beam of the 12m ALMA array. The synthesised beam is shown at
	the bottom left. The rectangle shows the slit position for the X-shooter
	observations, with a size of 1.2$\arcsec$$\times$11$\arcsec$.  The image
	is centered at the ALMA continuum peak,
	$\alpha$(J2000)\,=\,21$\rm^h$42$\rm^m$42$\rm^s$.631 and
	$\delta$(J2000)\,=\,$-$44$\rm^o$30$\arcmin$09$\arcsec$.44. Primary beam
	correction is not applied to the 1.2mm continuum image for clarity.
	North is up and east to the left.
	{\bf Middle: } Zoom in on the central region of interest.
	{\bf Top right: } X-shooter 2D \ly\, spectra along the slit as indicated
	in the left panel. To improve the signal-to-noise ratio in the
	spectrum, four contiguous piexels (4$\times$0.16$\arcsec$) were
	smoothed along the slit. The blue and red horizon-dashed
	lines denote the positions of the ALMA 1.2mm continuum and the \ly\,
	peak, respectively. {\bf Bottom right: } X-shooter extracted \ly\,
	spectra at the position of the ALMA 1.2mm continuum in blue and at the
	\ly\, peak in red.  Note that the velocity is relative to a redshift of
	2.3691.}
\label{cont}
\end{figure}

\begin{figure}[t]
\centering
\includegraphics[angle=0,width=0.75\textwidth]{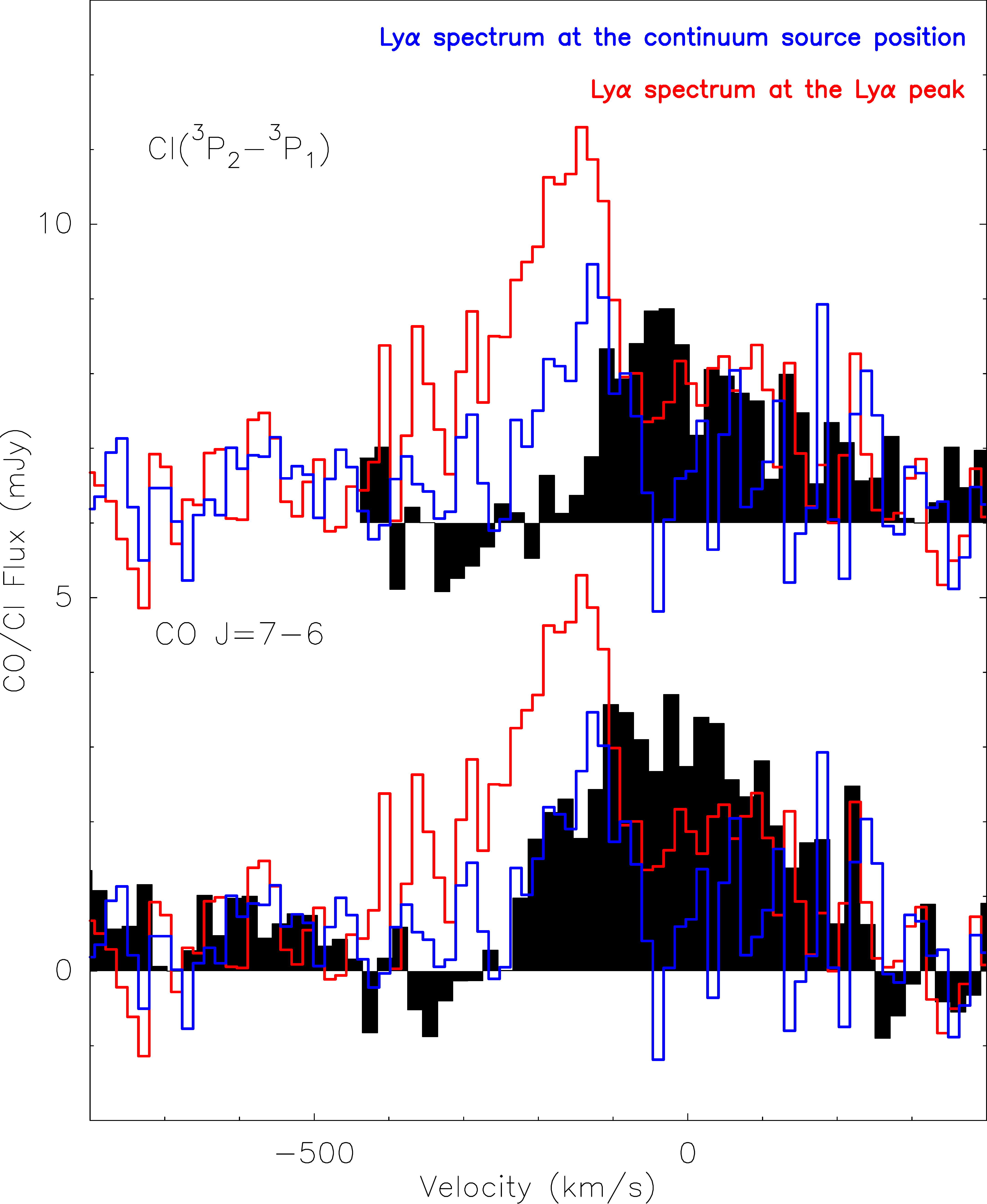}
\caption{Observed \co\, and \ctwo\, line profiles obtained with ALMA towards
	LAB6.1 are presented by filled histograms. X-shooter \ly\, spectra at
	the position of the ALMA 1.2mm continuum and at the peak of \ly\, emission
	are shown in blue and red, respectively. The velocity
	scale refers to that of the \co\, line at redshift $z$\,=\,2.3691.
	Note that \co\, and \ctwo, in units of mJy, are shown in the same panel
	but the latter with an offset along the y-axis.  \ly\, flux is on an
	arbitrary scale.}
\label{profile} 
\end{figure}

We used the Atacama Large Millimeter/Submillimeter Array (ALMA) to search for
\co\, and \ctwo\, line emission as well as for associated continuum emission
toward one LAB, LAB6, in the Francis cluster
(J2143$-$4423)\cite{Francis1996,Ao2015,Palunas2004}. Its \ly\,
emission is distributed over a region of $\sim$120 kpc, with a luminosity of
$\sim$5.2$\times$~10$^{\rm 43}$~${\rm erg~s^{-1}}$. The field of view of ALMA
at the central observed frequency of 246.6 GHz is about 24$\arcsec$, or 200~kpc
(physical) projected on the sky at z\,=\,2.37. LAB6 is significantly detected
in the 1.2~mm dust continuum with ALMA, as shown in Figure~\ref{cont}. The
continuum source, LAB6.1, with a flux density of 1.57\ppm0.11~mJy is located at
the center of the LAB, with an apparent size of
0.82$\arcsec\times$0.70$\arcsec$ (6.8~kpc$\times$5.8~kpc). It is marginally
resolved with a deconvolved size of 0.52$\arcsec\times$0.46$\arcsec$
(4.3~kpc$\times$3.8~kpc).  To highlight the gas kinematics in the central
galaxy, we show the line profiles of \co\, and \ctwo\, towards LAB6.1 in
Figure~\ref{profile}.  The \co\, line indicates a redshift of 2.3691\ppm0.0004.
In Figure~\ref{cont}, there is another unresolved continuum source, LAB6.2,
detected with a flux density of 1.60\ppm0.24~mJy. It is offset by
$\sim$16$\arcsec$ (130~kpc) from the center, and may be located outside the
LAB. Thus, LAB6.2 is unlikely to be physically associated with LAB6, which is
also supported by the evidence that no \co\, or \ctwo\, line emission is
detected from LAB6.2.

Our ALMA observations show the distribution and kinematics of gas from the
central galaxy. Figure~\ref{cont} shows that the \ly\,
emission 
is clearly peaked around the location of the
galaxy and is featured with extended morphology in the southeast. Does the
central SMG power the physically extended \ly\, emission? To answer this
question, we used X-shooter on Unit Telescope 2 of the Very Large Telescope
(VLT) at the European Southern Observatory (ESO) to conduct follow-up
spectroscopic observations to measure the line profile of \ly. A slit was
placed along LAB6 as shown in the left panel of Figure~\ref{cont} and the
results are presented in the right panel of Figure~\ref{cont}. The \ly\, line
emission is clearly detected around LAB6.1 with its peak approximately
$-$150~\kms\, off the systemic velocity obtained from \co. Interestingly,
different from other LABs, the profile of the observed \ly\, line appears to be
blue-skewed.

Given the resonant nature of the scattering cross-section, high near the line
center and low in the line wings, \ly\, photons usually escape from a static
cloud with a double-peaked profile\cite{Zheng2002}. With outflowing (inflowing)
neutral gas, the Doppler effect would lead to an enhanced red (blue)
peak\cite{Zheng2002,Dijkstra2006,Verhamme2006,Faucher2010}.  In
Figure~\ref{profile}, we overlay the line profile of \ly\, on those of \co\,
and \ctwo, showing an offset of $\sim$ $-$150\,\kms. It is very rare to
observe a blue-skewed \ly\, line profile from ionised gas in comparison with
the systemic velocity of molecular or atomic gas in an LAB. Such line profiles
can be well reproduced in cold stream
models\cite{Zheng2002,Dijkstra2006,Verhamme2006,Faucher2010}, while SF-based
models with outflows driven by mechanical feedback generally give rise to
characteristic red-skewed asymmetric profiles.  However, SF-based models with
infalling gas may also produce the observed line profile.  When a contracting
shell model (see details in Methods) is used to fit the observed \ly\, spectrum
and the results are shown in Figure~\ref{shell}, the inferred velocity of about
$-$34~\kms\, of the shell suggests that the infalling gas may become stalled
near the galaxy. For the 1.2\,mm continuum source detected by ALMA, LAB6.1, the
inferred far-infrared (FIR) luminosity is
(6.5$\pm$0.13)$\times$10$^{12}$~\lsol\, and its corresponding SFR is
approximately 1100 \msol\,yr$^{-1}$. Such a high degree of SF can easily
provide enough ionizing ultraviolet (UV) photons to produce extended \ly\,
emission if they escape.  Therefore, the observed \ly\, emission with
blue-skewed line profile can correspond to that either from the cooling
radiation or from a central photoionising source but scattered by the infalling
gas. In both scenarios, infalling gas is needed to produce the observed line
profiles presented in this paper.  As a whole, our observational result
represents direct kinematic evidence that gas is falling into the central
region where the SMG is located, but it does not necessarily imply cooling
radiation as the powering mechanism of the LAB.  We note that the gas infall
rate is approximately 1.7~\msol\,yr$^{-1}$ (see details in Methods), which is
negligible in comparison to the consumed material of approximately 1100
\msol\,yr$^{-1}$ by star formation.  The spatial offset ($\sim$5.4~kpc) between
the \ly\, peak and the dust continuum peak of LAB6 may be a result of
anisotropic gas and dust distribution and may be orientation dependent, as seen
in simulations \cite{Geach2016}.

Recently proposed SF-based models\cite{Cen2013} predict that LABs at high
redshift may correspond to protoclusters containing the most
massive galaxies and cluster halos in the early universe as well as ubiquitous
strong infrared sources undergoing extreme starbursts. These models also
predict that the \ly\, emission from photons that escape from a galaxy is
expected to be significantly polarised. This has been confirmed for the first
time for one LAB in the SSA22 field\cite{Hayes2011} and the detection of
polarised radiation is inconsistent with in situ production of \ly\, photons,
suggesting that it must have been produced in the galaxies hosted within the
nebula, and re-scattered by neutral hydrogen. Of course, the galaxies already
formed in the LABs will likely contribute to the extended \ly\, emission with
photoionization by young stars and/or AGNs. Stacking of faint LAEs in the SSA
22 field\cite{Matsuda2012} shows extended and faint \ly\, emission surrounding
the bright sources at the center, implying possible connection of the extended
\ly\, emission with the central powering source.

For the scenario of cold streams, the radiatively cooled gas can continuously
fall onto the center of the LAB to form a new galaxy or to feed the existing
central galaxy.  The gas inflow rate may be high enough to feed the central
starburst and significantly grow the galaxy, as shown in
simulations\cite{Cen2014,Dekel2009,Agertz2011}. This is in apparent contrast
with our inferred low accretion rate from the contracting shell model.  Note
that the mophology and the three-dimensional distribution of cold
streams\cite{Dekel2009,Agertz2011} are far from a spherical shell structure.
While the shell model result supports the cold stream scenario, details need to
be scrutinized and realistic models are necessary. It is likely that the cold
accretion streams are connected to large-scale filaments. Observational
evidence includes the kinematics of gas distribution of cosmic web near a
quasar\cite{Martin2015} and a 12~Mpc (comoving) filamentary structure traced by
LABs in an overdense region\cite{Erb2011}. Given such a picture, LABs could be
a type of objects that host a diverse sample of galaxies, or in the extreme
case are devoid of galaxies.  In the latter case, LABs are part of the cosmic
web in which galaxies have not yet formed or still at their early stage.  Our
recent deep observations\cite{Ao2017} revealed that the central heating sources
are still missing among about two third of LABs in the SSA 22 field. It may be
caused by the poor sensitivities of these observations or the fact that the
extended \ly\, emission is dominated by cooling radiation in some LABs.

For the origins of the infalling gas, besides the
long-sought-after cold accretion streams, it is possible that the gas was
ejected earlier as part of the galactic wind from the central region and is
falling back (the so-called galactic fountain). Comparing to the cold
accretion streams, gas in the latter case is expected to have larger covering
factor.  If cold streams or infalling of recycled gas are ubiqutious,
blue-skewed \ly\, line profiles should not be so rare. However, the gas
distribution can become more complex with feedbacks from central sources and
the effect on \ly\, emission can depend on viewing directions\cite{Zheng2014},
which may help explain the rarity of the blue-skewed prolies in the observed
\ly\, spectra. To constrain the origin of the infalling gas, it would be useful
to study LABs without strong feedbacks from central sources with the help of
deep ALMA observations. Observations that provide constraints on the properties
of the gas (e.g. metallicity and column density) and \ly\, radiative transfer
modeling with realistic gas distribution from galaxy formation simulations can
further disentangle different scenarios of its origin, improving our
understanding of galaxy formation in the early Universe.

\clearpage
{\noindent \large \bf Acknowledgements} Y.A. acknowledges financial support by
NSFC grant 11373007. J.E.G. is supported by a Royal Society University
Research Fellowship. This paper makes use of the following ALMA data:
ADS/JAO.ALMA\#2015.1.00952.S. ALMA is a partnership of ESO (representing its
member states), NSF (USA) and NINS (Japan), together with NRC (Canada), NSC and
ASIAA (Taiwan), and KASI (Republic of Korea), in cooperation with the Republic
of Chile. The Joint ALMA Observatory is operated by ESO, AUI/NRAO and NAOJ.
Based on observations collected at the European Organisation for Astronomical
Research in the Southern Hemisphere under ESO programme IDs 297.A-5059(A) and
082.A-0846(B).  This research has made use of the SVO Filter Profile Service
(http://svo2.cab.inta-csic.es/theory/fps/) supported from the Spanish MINECO
through grant AyA2014-55216.

{\noindent \large \bf Author Contributions} Y.A. is the Principal Investigator
of the ALMA and VLT/X-shooter observing proposals. Y.A. reduced the data and
wrote the initial manuscript. Z.Z. conducted the data analysis with the SED
modeling, drafted the interpretation and discussion of the \ly\, emission, and 
polished the manuscript, and
S.N. performed the \ly\, spectrum fitting with an infalling shell model.  C.H.
helped to polish the manuscript. All authors discussed and commented on the
manuscript.

{\noindent \large \bf Competing Interests} The authors declare no competing
interests

{\noindent \large \bf Additional information}\\ {\noindent \large \bf Extended
data} is available in the online version of the paper.\\ {\noindent \large \bf
Correspondence and requests} for materials should be addressed to Y.A. (ypao@pmo.ac.cn).

\clearpage

{\noindent \bf \Large Methods}
\setcounter{figure}{0}
\setcounter{table}{0}
\renewcommand{\thefigure}{S\arabic{figure}}
\renewcommand{\thetable}{S\arabic{table}}

{\noindent \bf \large ALMA observations} 
We observed LAB6 in the protocluster J2143-4423\cite{Ao2015}.  The observations
were carried out on 2016 May 7th with 36 12 meter antennas and baselines
between 16.5 and 629~m. The total on-source observing time was 25 minutes. We
used the correlators in the Frequency Division Mode (FDM) at the central
frequency of 238.76 GHz in the lower sideband and in the Time Division Mode
(TDM) at the central frequency of 254.51 GHz in the upper sideband. The data
were reduced with the Common Astronomy Software Application (CASA) package in a
standard manner.  Originally, Pallas was adopted for the flux calibrator. The
Pallas-based flux calibration resulted in a flux density of 1.17 Jy at 253.5
GHz for the bandpass calibrator J2056-4714. This is 1.28 times the flux density
based on the ALMA calibrator catalog, estimated by interpolation between data
entries in band 3 (97.5~GHz) and 7 (343.5~GHz). In the ALMA calibrator catalog,
there is no signature of high variability for J2056-4714 around the observing
date.  The flux discrepancy may be attributed to the Pallas-based flux
calibration.  Therefore, the flux density of the phase/gain calibrator
J2139-4235 has been re-scaled using the flux of the bandpass calibrator
interpolated with the measurements in the ALMA catalog. The primary beam size
is about 24$\arcsec$ at the central frequency of 246.6 GHz. The achieved
synthesised beamsize (full-width at half maximum, FWHM) is
0.64$\arcsec\times$0.51$\arcsec$ with a position angle of 9 degrees east of north. The
continuum maps have an rms noise level of 40~$\mu$Jy/beam.  The flux densities
of the detected sources were measured with the CASA task IMFIT.

\begin{figure}[h]
	\centering
	\includegraphics[angle=0,width=1\textwidth]{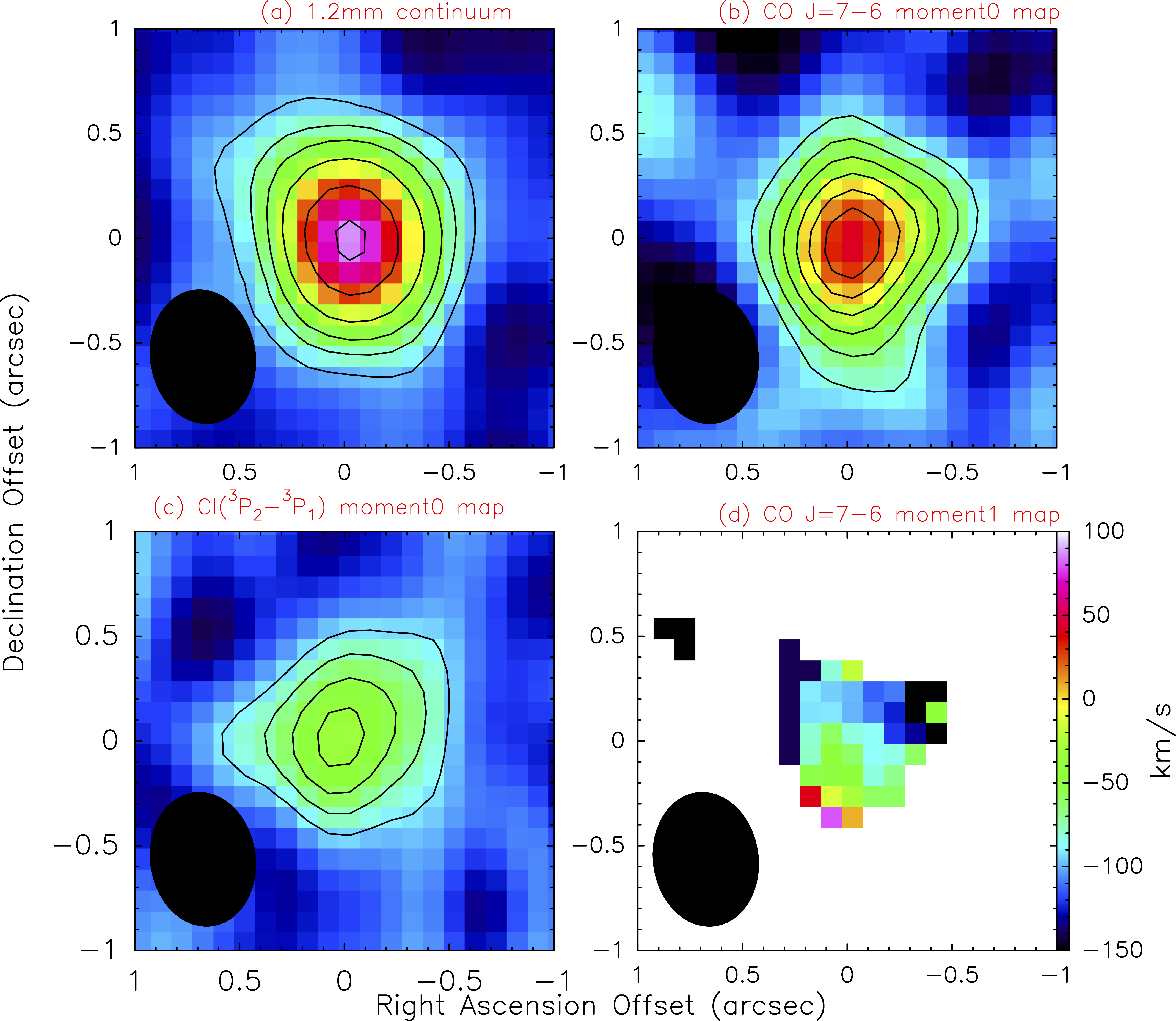}
	\caption{Maps of LAB6.1 revealed by ALMA. a). The 1.2mm continuum in
	color scale and contours. Contours are 3, 5, 7, 10, 15, 20
	$\times$ 0.045 mJy~beam$^{-1}$. b). The \co\, integrated intensity in
	color scale and contours.  Contours are 3, 5, 7, 9, 11, 13
	$\times$ 0.075 Jy~\kms~beam$^{-1}$.  c). The \ctwo\, integrated
	intensity in color scale and contours.  Contours are 3, 5, 7, 9
	$\times$ 0.06 Jy~\kms~beam$^{-1}$.  d).  \co\, velocity field relative
	to a redshift of z\,=\,2.3691 on a color scale. The synthesised beam of
	0.82$\arcsec\times$0.70$\arcsec$ is shown in the lower left corner of
each panel.} \label{b6} 
\end{figure} 
Two lines are detected towards LAB6.1 (see Figure~\ref{profile}). Accounting
for the known offset between the frequencies of the \co\, and \ctwo\, lines,
one can estimate an accurate redshift of 2.3796\ppm0.0004 for LAB6.1.  This is
within the uncertainties consistent with the redshift of the \ly\, emission of
LAB6. Figure~\ref{b6} shows a $\lambda$ $\sim$ 1.2 mm continuum map, integrated
intensity maps of \co\, and \ctwo\, as well as the velocity field traced by
\co\, towards LAB6.1.  In Figure~\ref{b6}(d), the projected velocity varies by
about $200 {\rm km\, s^{-1}}$ across $\sim 0.7\arcsec$ ($\sim 6 {\rm kpc}$).
It can be a hint of rotation of the molecular gas in the central galaxy.  The
other possibility is gas inflow or outflow. We note that the gradient of the
velocity field approximately coincides with the direction towards the diffuse
\ly\, emission to the southeast (Fig.~\ref{cont}).

{\noindent \large \bf VLT/X-shooter observations} We obtained a UV to NIR
spectrum (project ID: 095.A-0764) at the Very Large Telescope (VLT) with the
VLT/X-shooter instrument\cite{Vernet2011} from Oct.  20 to 26, 2016.  The
X-shooter instrument consists of three spectroscopic arms: UVB, VIS and NIR,
ranging from 293.6 nm to 2480 nm. The observations were taken with
1.6$\arcsec$$\times$11$\arcsec$, 1.5$\arcsec$$\times$11$\arcsec$, and
1.2$\arcsec$$\times$11$\arcsec$ slits for the three arms, respectively.  The
total on-source integration time is 86 minutes. The data
reduction has been performed using the {\bf reflex} X-shooter pipeline
2.9.1\cite{Freudling2013}.

\begin{figure}[h] \centering
	\includegraphics[angle=0,width=1\textwidth]{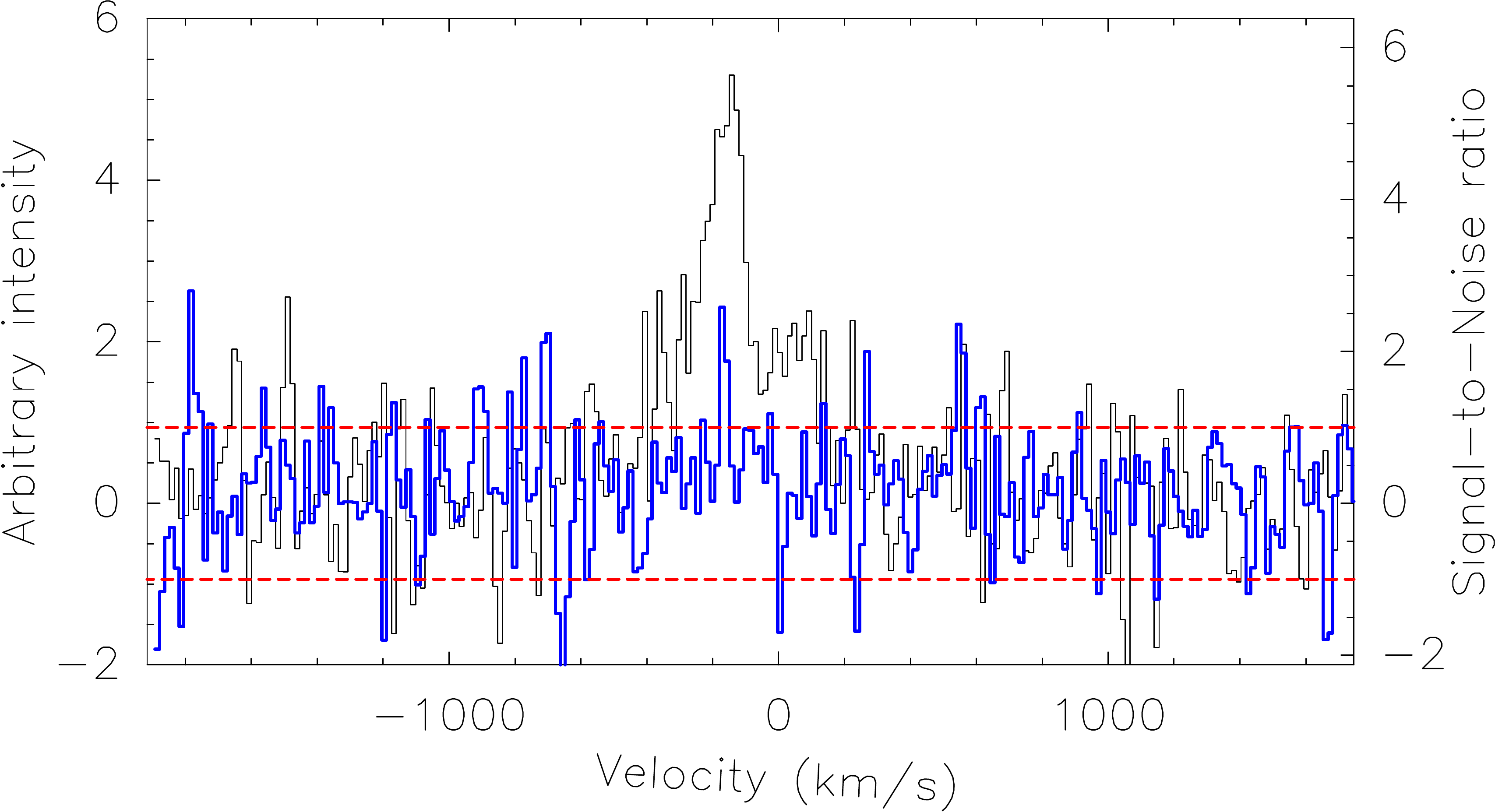}
	\caption{X-shooter extracted \ly\, spectra at the \ly\, peak in black
	and a selected position in the southeastern part in blue, where there is
	extended \ly\, emission. The 1$\sigma$ range of the noise
	is shown in red dashed lines and signal-to-noise ratios are labelled in the right
	axis. Note that the velocity is relative to a redshift of 2.3691.} 
\label{spec_se} \end{figure}

Note that only the UVB arm covering the \ly\, spectroscopic data is presented
in this paper. To improve the signal to noise ratio of the spectrum, four
contiguous pixels (4$\times$1.6$\arcsec$) were smoothed along the slit. To
check the \ly\, line emission in the southeast part (see the left panel of
Figure~\ref{cont} for this extended emission), we present a typical spectrum from this
region in Figure~\ref{spec_se}, where the line emission is not detected.
Apparently this region is too faint for our X-shooter observations.
For the spectrum at the \ly\, peak, the blue peak is obvious
and the signal-to-noise ratios (SNRs) at 7 channels are over 4. For the red peak,
there are 9 continuous channels with SNRs around 2, and the SNR will become 6
by smoothing all channels.
The \ly\, luminosity around LAB6.1 recovered by X-shooter is
$\sim$5.0$\times$10$^{\rm 42}$~${\rm erg~s^{-1}}$, which is about 10\% of
the total luminosity in the whole LAB\cite{Palunas2004}. In combination with 
the data at other bands in the next section, the \ly\, escape fraction is
estimated to be only $\sim$5\%.

Except for the \ly\, emission, there is no significant
intestellar absorption lines are detected among the three arms.

\begin{figure}[h] \centering
	\includegraphics[angle=0,width=1\textwidth]{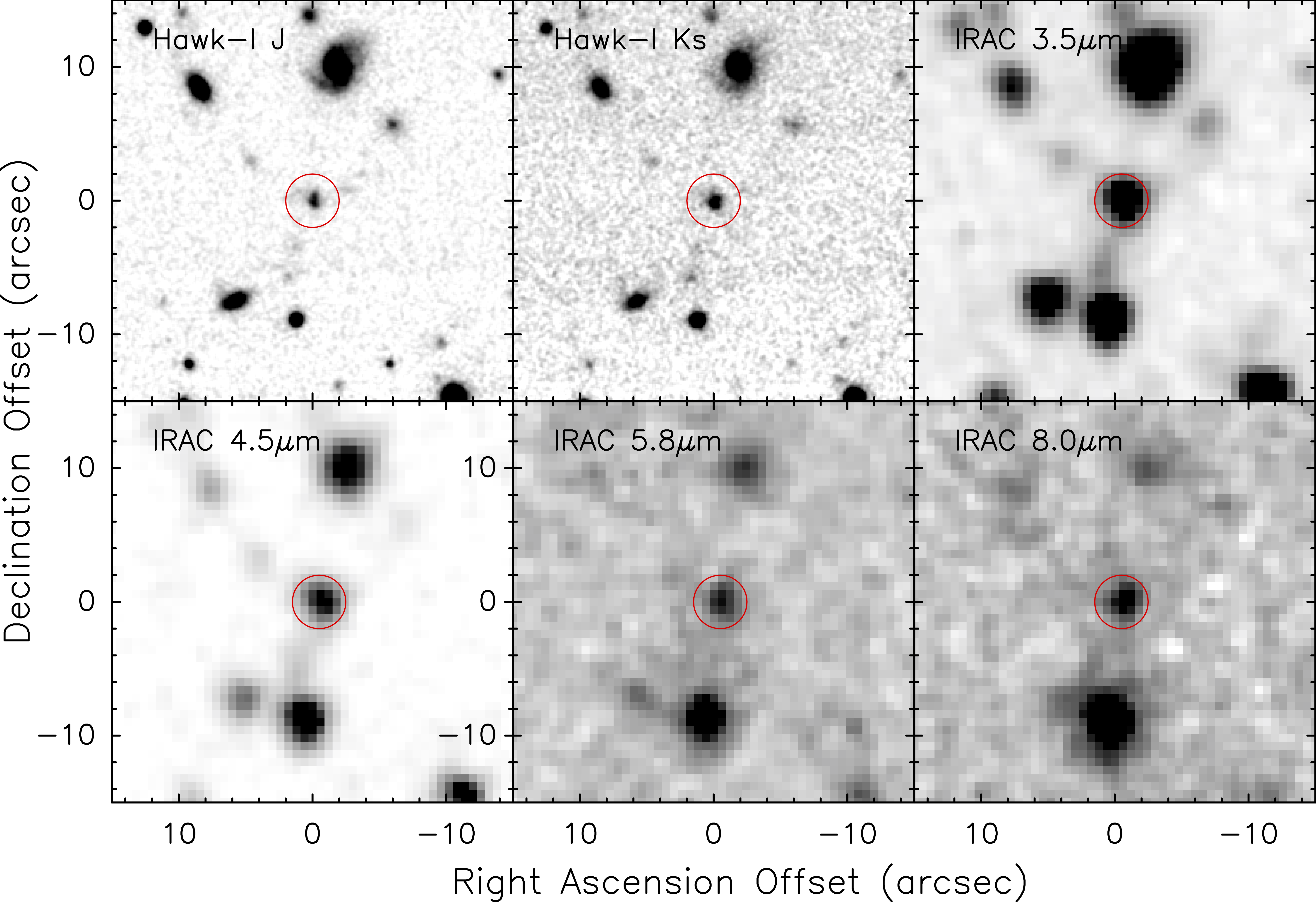}
	\caption{Multi-wavelength images of LAB6.1 from VLT HAWK-I and {\it
Spitzer} IRAC observations. The field of view is
30$\arcsec$\,$\times$\,30$\arcsec$.} \label{images} \end{figure}

{\noindent \large \bf Near- to mid-infrared data and constraints on the central
source}\label{sed} To constrain the physical properties of the central galaxies
in the LABs of the Francis cluster J2143-4423, {\it Herschel} PACS and SPIRE
data have been used for data analysis\cite{Ao2015}. However, the coarse
angular resolution may induce large uncertainties as shown in Figure~\ref{cont}
where two sources with a separation of only 16$\arcsec$ are revealed by ALMA.
Higher angular resolution is therefore required to constrain the central
sources. 

Near-infrared data were taken with the High Acuity Wide field K-band Imager
(HAWK-I) on VLT during October and November 2008. Broadband filters, J band at
1.258~$\mu$m and Ks band at 2.146~$\mu$m, were used for the observations.
Mid-infrared data were obtained using the {\it
Spitzer} Infrared Array Camera (IRAC) at 3.6,
4.5, 5.8 and 8.0~$\mu$m\cite{Colbert2011}. The details of the data reductions
can be found in the literature\cite{Colbert2011}. Data are listed in
Table~\ref{tab_photo} and images are presented in Figure~\ref{images}.

Using Bayesian Analysis of Galaxies for Physical Inference and Parameter
EStimation (Bagpipes)\cite{Carnall2018,Rodrigo1,Rodrigo2}, we constrain the
physical properties of the host galaxy from SED modeling with near- to
mid-infrared data.  If the redshift is left as a free parameter, the
photometric redshift derived from Bagpipes is 2.53$_{-0.38}^{+0.33}$,
consistent with that determined from the submm lines.  To constrain the
physical properties, we fixed the redshift to 2.3691 and found that the SED can
be well described by a recent burst with an SFR of
1.18$^{+0.39}_{-0.56}\times10^3$ \msol~yr$^{-1}$, a stellar age of
62$^{+37}_{-22}$ Myr and a logarithmic stellar mass log(M$_{\rm
stellar}$/\msol) = 11.00$^{+0.08}_{-0.14}$, as shown in Figure~\ref{sed}.

\begin{center} 
\begin{table*}[t] 
\centering \caption{Photometric data towards LAB6.1}\label{tab_photo}
\begin{tabular}{ccccccc} 
\hline
	Bands  & HAWK-I J  &  HAWK-I K & IRAC1  & IRAC2 & IRAC3 & IRAC4   \\
\hline
	$\lambda$ ($\mu$m) & 1.258 & 2.146 & 3.6 & 4.5 & 5.8 & 8.0  \\
        flux ($\mu$Jy)$^a$  & 2.56\ppm0.06 & 9.01\ppm 0.26 & 24.79\ppm0.17 & 
	30.953\ppm0.26 & 40.29\ppm0.95 & 43.53\ppm1.60 \\
\hline 
\end{tabular} 
\begin{list}{}{}
\item{$^{\mathrm{a}}$ For the SED modelling, a 15\% error bar is adopted for
	each IRAC band to account for the calibration uncertainty.}
\end{list}
\end{table*}
\end{center}

\begin{figure}[h] \centering
	\includegraphics[angle=0,width=1\textwidth]{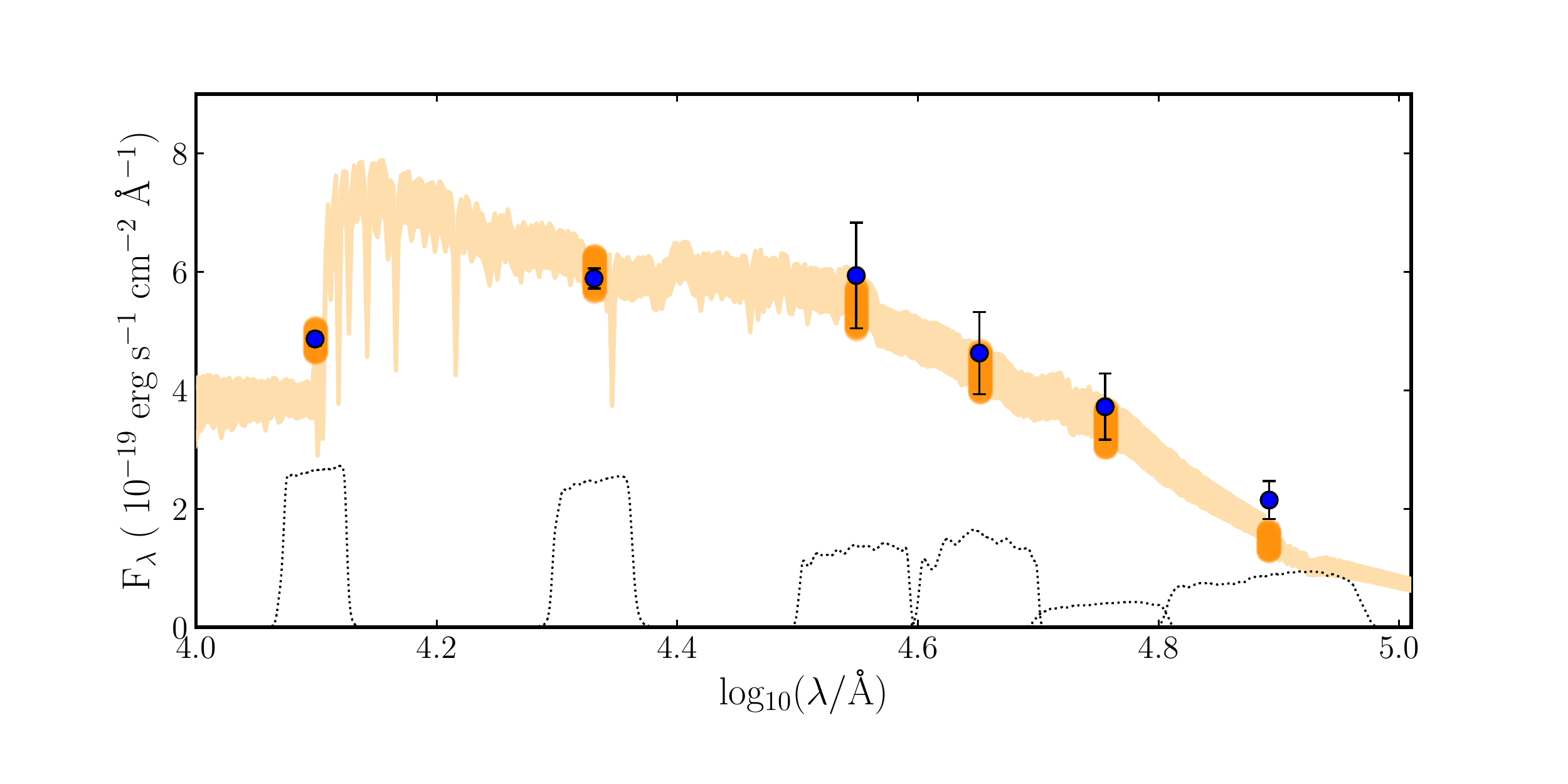}
	\caption{SED modelling result for LAB6.1. Measurements in the observed
	near- and mid-infrared bands are shown in blue. The 16th to 84th
	percentile range for the posterior spectrum is shaded in light orange and that of 
        the photometry in dark orange (at the observed wavelengths). 
        Dotted curves correpond
	to the filter transmissions.}\label{sed} 
\end{figure}

{\noindent \large \bf Molecular gas mass, dust mass and star formation rate}
Here we derive the physical properties of LAB6.1 using \co\, and dust emission.
For LAB6.1, the measured flux density at 1.2mm is 1.57\ppm0.11~mJy, which is
significantly lower than the predicted value of 2.43 mJy by the SED model of
LAB6\cite{Ao2015}. This latter higher value may be due to the contribution
from other spatial components. As shown in Figure~\ref{cont}, there is a
continuum source located 16$\arcsec$ from LAB6.1. In the {\it Herschel} SPIRE
bands used for the SED model\cite{Ao2015}, this source is blended with LAB6.1,
which may explain the overprediction of the 1.2mm flux density from the
previous SED model.  Assuming that the total observed FIR
luminosity$\cite{Ao2015}$ originates from both sources and the fractional
contribution from LAB6.1 follows that of the 1.2~mm continuum, we derived the
FIR luminosity of LAB6.1 to be 6.5$\pm$0.13$\times$10$^{12}$~\lsol, which
translates to an SFR of 960 \msol~yr$^{-1}$ based on the SFR-L$_{\rm FIR}$
relation\cite{Kennicutt2012}. 

With ALMA, the measured \co\, integrated flux density is 1.14$\pm$0.12
Jy\,\kms\,  and the corresponding line luminosity is
1.10$\pm$0.12$\times$10$^{8}$~\lsol\, or
6.50$\pm$0.71$\times$10$^{9}$~K~\kms\,pc$^2$. Following the SFR calibration
using the \co\, line luminosity\cite{Lu2015}, the derived SFR is 1240
\msol~yr$^{-1}$.  From the mean value of the SFRs from the above two methods,
the derived SFR of LAB6.1 is approximately 1100 \msol~yr$^{-1}$, which
is consistent with the value derived with the near- and mid-infrared SED
fitting.

Assuming a line ratio of 0.24 for $L_{\rm CO(7-6)}$/$L_{\rm
CO(1-0)}$\cite{Carilli2013,Yang2017} and the standard factor
$X$$\rm_{CO}$\,=\,0.8 M$\rm_\odot$ $\rm(K\ \kms\ pc^2)^{-1}$ for ultra-luminous
infrared galaxies\cite{Downes1998}, we find a molecular gas mass of
(2.2$\pm$1.1)$\times$10$^{10}$\,M$\rm_\odot$. The dust mass from the SED
model\cite{Ao2015} is (2.1$\pm$0.5)$\times$10$^{8}$\,M$\rm_\odot$ after
rescaling the result according to the continuum flux density measured by
ALMA.  This implies a gas to dust mass ratio of 105$\pm$60.

{\noindent \bf \large Shell model fit to the \ly\, spectrum} 
The \ly\, line emergent from \ly\, emitting galaxies and LABs is usually
characterised by a profile skewed toward the red side of the systemic
velocity\cite{Yang2014}, which can be understood as \ly\, emission escaping
from outflowing gas after experiencing resonant scatterings off neutral
hydrogen atoms.  The resonant scattering cross-section is high near the line center
and low at wings, and \ly\, emission from a static cloud (e.g. with no inflow
or outflow) would have frequencies shifted away from the line center to escape,
leading to a double-peaked line profile\cite{Zheng2002}. For an expanding cloud
(with outflowing gas), the Doppler effect makes \ly\, photons back scattered
from the far side of the cloud easier to escape from the near side. This
results in a red-skewed \ly\, line profile\cite{Zheng2002,Verhamme2006}, producing 
a \ly\, line with an enhanced red peak and a suppressed blue peak.
However, the \ly\, spectrum of LAB6.1 shows a blue-skewed line profile, with an
enhanced blue peak and a suppressed red peak. Such spectral features indicate
gas infall.

For LAB6.1, there is a slight offset between the peak of the \ly\, emission and
the central galaxy associated with the 1.2mm dust continuum (see
Fig.~\ref{cont}), which is about 0.64$\arcsec$ (5.4kpc). This raises the
question about the nature of the gas probed by the \ly\, emission. If we take
the velocity ($\sim -50 {\rm km\, s^{-1}}$) near the trough between the red and
blue peaks in the \ly\, spectrum as an indication of the bulk velocity of the
gas (with respect to the continuum source), the gas can be either on the near
side outflowing from the galaxy or on the far side inflowing toward the galaxy.
However, the blue-skewed \ly\, spectrum suggests that the gas itself is
collapsing. Furthermore, the \ly\, spectrum centered on the galaxy has a
similar shape to that near the peak emission.  The shape of the spectrum at the
peak \ly\, emission and that at the position of the central galaxy provide
strong support that the gas is falling into the central galaxy. The small
spatial offset of the \ly\, emission from the galaxy may be caused by 
the non-uniform distribution of the gas, e.g. accretion along filaments,
or by a significant absorption near the continuum source, leaving the dominant
contribution to the surviving \ly\, emission biased toward the low-density 
accretion flows.

As an attempt to model the \ly\, spectrum, we adopt a simple model that
infalling gas is uniformly located in a thin shell of neutral gas and that the
\ly\, radiation from the central source is scattered by the hyrogen atoms in
the shell. This is the reverse version of the expanding shell
model\cite{Verhamme2006} commonly used to model the \ly\, spectra from
star-forming galaxies. We characterise the model with three parameters, the
infalling velocity $V$ of the shell and the temperature $T$ and column density
$N_{\rm HI}$ of the neutral hydrogen in the shell. A Markov Chain Monte Carlo calculation
is performed to constrain the parameters, based on interpolating a compilation
of \ly\, spectra\cite{Gronke2015} from the shell model.  We find that the best
fit (model 1) has $V=-33.8^{+3.2}_{-3.1} {\rm \,km\, s^{-1}}$, $\log(T/{\rm
K})=3.60^{+0.27}_{-0.27}$, and $\log(N_{\rm HI}/{\rm
cm}^{-2})=19.80^{+0.05}_{-0.05}$, with $\chi^2 =53.5$ for 74 degrees of
freedom. The $\chi^2$ value is acceptable for 74 degrees of freedom
(which has 1$\sigma$ range of $\chi^2$ values of 74$\pm$12), and it does not
imply over-fitting.
The other acceptable but much less favored solution (model 2) with
$\chi^2=68.9$ has $V=-93.9^{+20.1}_{-14.7} {\rm \,km\, s^{-1}}$, $\log(T/{\rm
K})=4.19^{+0.22}_{-0.19}$, and $\log(N_{\rm HI}/{\rm
cm}^{-2})=19.62^{+0.05}_{-0.05}$.  The two model fit spectra are shown in
Figure~\ref{shell}. While the first model clearly works better, it misses the
\ly\, flux near $v=0$ in the red peak, which implies that a uniform shell
model is over-simplified. The presence of \ly\, flux between two peaks suggests
possible contribution to the \ly\, emission of extended origin (i.e. cooling,
photoionization, etc).  The crude picture from the model is that the infalling
gas (with velocity of $\sim 34\,{\rm km\, s^{-1}}$) has a temperature of $\sim
4000$K and neutral hydrogen column density of $10^{19.8}{\rm cm}^{-2}$ (in the
regime of the Lyman-limit system). Given the spatial extent of the major \ly\,
emission in the 2D spectrum, which is about 1$\arcsec$ in radius ($\sim$8~kpc
on a physical scale) and the offset from the central galaxy ($\sim$ 5.4~kpc),
the gas can be regarded as part of the circumgalactic medium. 

There could be two possible origins of the infalling gas. First, it may come
from the long-sought-after cold streams. The infalling gas may be getting
stalled near the central galaxy (hence the low velocity). The other scenario
may be that the gas was ejected earlier as part of the galactic wind from the
central region and is falling back (the so-called galactic fountain).  If we
take $r\sim 8{\rm kpc}$ as the radius of the shell, the total amount of
hydrogen mass is then $4\pi r^2 N_{\rm HI} m_H \sim 4.0\times 10^8 \msol$.
The dynamical time scale is $R/|V|\sim 230\, {\rm Myr}$ for $|V|\sim 34\, {\rm km\,
s^{-1}}$. The gas inflow rate is then about $1.7\msol\, {\rm yr}^{-1}$. This
suggests that while the infalling gas is important in shaping the \ly\,
emission, it is not the main component responsible for the growth of the
central galaxy. The shell model is overly simplified, and the gas distribution
can be more complex.  
For example, the inferred SFR in the power source should produce powerful outflows,
which in turn alter the gas kinematics even in the CGM. The photoionisation of the
CGM gas by ionising photons from the central source can also change the distribution
of the neutral gas as well as produce \ly\, emission. 
The above modelling result should be taken as a rough estimate.  \ly\, radiative transfer
calculation with galaxies and gas distribution in hydrodynamic galaxy formation simulations 
would lead to better insights into systems like LAB6.1.

In a previous work, one $z\sim 2.3$ LAB system, CDFS-LAB10\cite{Yang2014}, was
found to have a \ly\, component that is blueshifted (about $-500 {\rm km\,
s^{-1}}$) with respect to a reference galaxy. However, this \ly\, component
itself is not associated with any galaxy, and is about 2$\arcsec$ (16.4~kpc)
away from the reference galaxy. Compared to LAB6.1, this component in
CDFS-LAB10 is much farther away from the galaxy, both in projected distance
(16.4~kpc versus 5.4~kpc) and in velocity ($\sim -500 {\rm km\, s^{-1}}$
versus $\sim -50 {\rm km\, s^{-1}}$). In addition, the \ly\, spectrum of the
component in CDFS-LAB10 appears to be double peaked with roughly symmetric red
and blue peaks, indicating that the gas itself is not contracting or expanding.
This is in direct contrast to the blue-skewed \ly\, profile of LAB6.1.
Therefore, given the large spatial and velocity separation of the \ly\,
component in CDFS-LAB10 from the reference galaxy and the roughly symmetric
double peaks, its relation to the galaxy is not clear, with infalling,
outflowing, or photoionised gas at large distances all remaining as
possibilities. \ly\, emission with prominent blue peaks is also detected in
two $z\sim 3.3$ LABs \cite{Vanzella2017}, and both show multiple emission
components in the images.  Also in both LABs, the \ly\, line profiles are
double-peaked, more symmetric that that of LAB6.1, making it hard to establish
a clear association with inflowing gas.  If we extend the case to \ly\,
emitting galaxies (rather than LABs), it is found that in a sample of 237 MUSE
\ly\, spectra about 5\% are well-fitted through a inflowing shell
\cite{Gronke2017}, suggesting the existence of infalling gas in a small
fraction of systems. 

Compared to previous detections of \ly\, spectra with blue peaks, the LAB6.1
case has more blue-skewed \ly\, line profile. It also has molecular/atomic gas
and dust detections. The peculiar \ly\, line profile and the rich data make
this object worth further investigations. For LAB6.1, the close association of
the  \ly\, emission in position with the central galaxy probed by the 1.2mm
continuum (see Fig.~\ref{cont}) and the more extended morphology suggest that
gas probed by the \ly\, emission surrounds the galaxy. The blue-skewed \ly\,
line can be produced by gas infall.  The reasonably good fit to the line
profile with a contracting shell model lends further support to such a
scenario. In summary, the \ly\, observation in combination with the ALMA
obervation of LAB6.1 provides the first clear case of blue-skewed \ly\, line in
an LAB system, which indicates gas infall.

\begin{figure}[h]
\centering
\includegraphics[angle=0,width=1\textwidth]{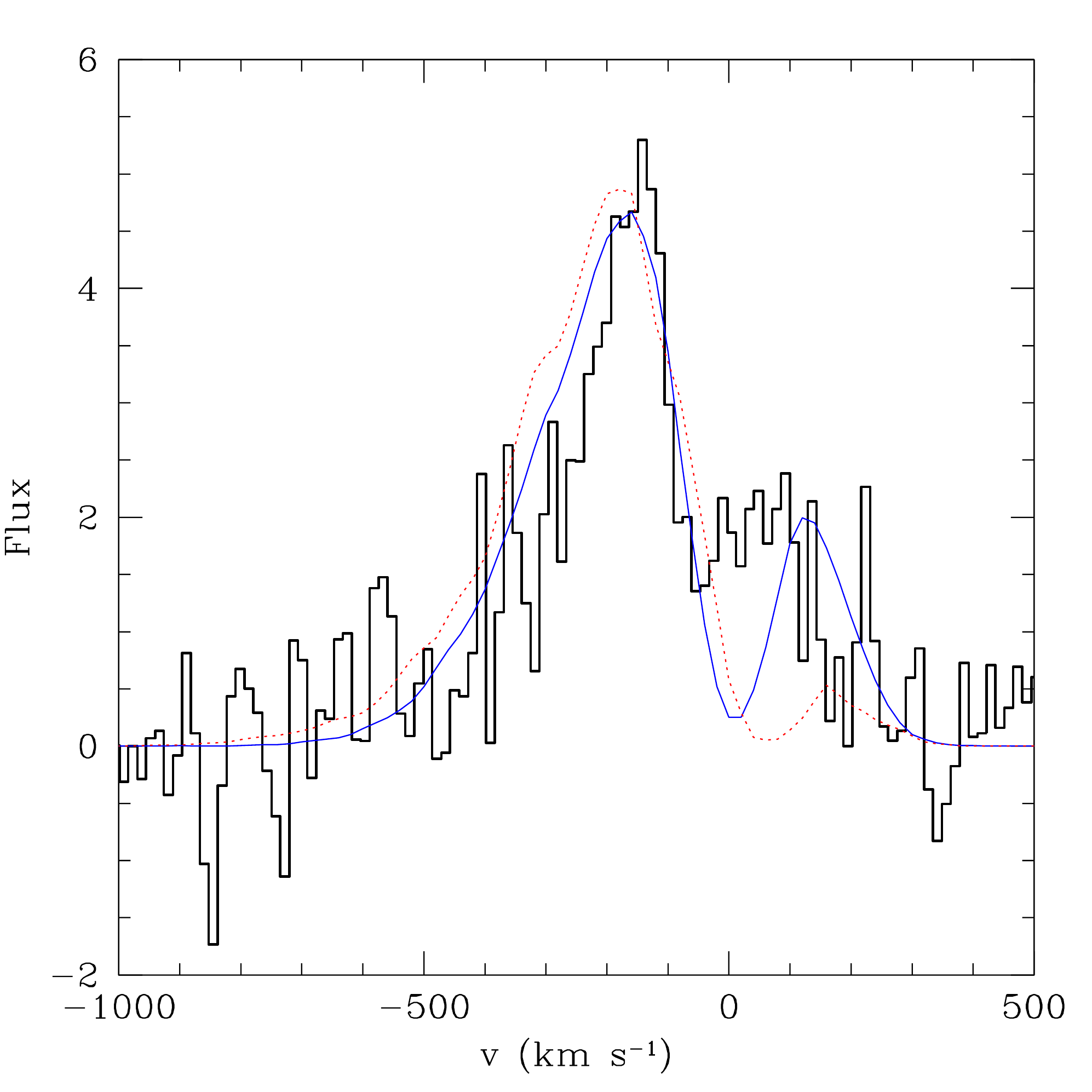}
	\caption{Shell model fits to the \ly\, emission (see details in
	Methods).  Two local minima shell model fits are shown as blue (model
	1) and red (model 2) curves, with model 1 the preferred fit  (see the
	text for details of the model parameters).}\label{shell} 
\end{figure}

\end{document}